\documentclass{article}

\PassOptionsToPackage{numbers, sort&compress}{natbib}
 \usepackage[preprint]{neurips_2025}

\usepackage[utf8]{inputenc} 
\usepackage[T1]{fontenc}    
\usepackage{url}            
\usepackage{booktabs}       
\usepackage{amsfonts}       
\usepackage{amsmath}
\usepackage{nicefrac}       
\usepackage{microtype}      
\usepackage{xcolor}         
\usepackage[colorinlistoftodos,prependcaption,textsize=tiny]{todonotes}
\usepackage{wrapfig}
\usepackage{siunitx}
\usepackage{amsthm}
\usepackage{cleveref}
\usepackage{braket}

\theoremstyle{plain} 
\newtheorem{theorem}{Theorem}
\newtheorem{proposition}[theorem]{Proposition}

\theoremstyle{remark} 

\theoremstyle{definition} 

\title{Generative Latent Space Dynamics of Electron Density}

\author{%
  Yuan Chiang\thanks{work started as a student at Lawrence Livermore National Laboratory.} \\
  University of California Berkeley\\ 
  Lawrence Berkeley National Laboratory \\
  \texttt{cyrusyc@berkeley.edu} \\
  \And
  Youngsoo Choi\\
  Center for Applied Scientific Computing\\
  Lawrence Livermore National Laboratory\\
  \texttt{choi15@llnl.edu} \\
  \And
  Daniel Osei-Kuffuor\\
  Center for Applied Scientific Computing\\
  Lawrence Livermore National Laboratory\\
  \texttt{oseikuffuor1@llnl.edu} \\
}

\begin{document}

\maketitle

\begin{abstract}
Modeling the time-dependent evolution of electron density is essential for understanding quantum mechanical behaviors of condensed matter and enabling predictive simulations in spectroscopy, photochemistry, and ultrafast science. Yet, while machine learning methods have advanced static density prediction, modeling its spatiotemporal dynamics remains largely unexplored. In this work, we introduce a generative framework that combines a 3D convolutional autoencoder with a latent diffusion model (LDM) to learn electron density trajectories from \textit{ab-initio} molecular dynamics (AIMD) simulations. Our method encodes electron densities into a compact latent space and predicts their future states by sampling from the learned conditional distribution, enabling stable long-horizon rollouts without drift or collapse. To preserve statistical fidelity, we incorporate a scaled Jensen-Shannon divergence regularization that aligns generated and reference density distributions. On AIMD trajectories of liquid lithium at 800 K, our model accurately captures both the spatial correlations and the log-normal-like statistical structure of the density. The proposed framework has the potential to accelerate the simulation of quantum dynamics and overcome key challenges faced by current spatiotemporal machine learning methods as surrogates of quantum mechanical simulators.
\end{abstract}

\section{Introduction}

The theoretical description of electrons and nuclei forms the cornerstone of understanding the physical and chemical properties of matter, yet it remains one of the most challenging frontiers of modern quantum mechanics. Electronic structure calculations, while offering a pathway to predict the ground and excited states of quantum many-body systems, are computationally expensive, with costs scaling steeply with the number of electrons.

In fact, full configuration interaction (FCI), although providing the exact solution to the time-independent Schrödinger equation, scales as $\mathcal{O}(n!)$ with respect to the number of molecular orbitals and basis set size. Coupled-cluster theory, while mitigating this issue through an exponential ansatz, still suffers from steep scaling: for a maximum excitation order $r$, the cost is $\mathcal{O}(n^{2r+2})$, where $n$ is the number of basis functions. The gold-standard CCSDT method (singles, doubles, and full triples excitations) scales as $\mathcal{O}(n^8)$, restricting its application to systems of at most tens to hundreds of atoms. Even Kohn–Sham density functional theory (KS-DFT), widely regarded as a computationally affordable alternative, typically scales as $\mathcal{O}(n^3)$, which becomes prohibitive for large-scale simulations or for repeated evaluations in dynamical settings. 

A promising strategy to accelerate such calculations is to provide a high-quality initial guess for the electron density or wavefunctions \cite{gubler2025accuracy,pulay1980convergence, broyden1965class, das2023accelerating}, which can significantly speed up convergence in self-consistent field (SCF) iterations. While most machine learning approaches to date have focused on predicting static electron densities from molecular geometries \cite{fu2024recipe, koker2024higher, jorgensen2022equivariant, jorgensen2020deepdft, achar2023machine, li2025image}, far fewer address the dynamical evolution of the electron density. However, this dynamical information is essential: the time-dependent electron density encodes rich physical observables, such as the dynamic structure factor, excitation energies, and transition moments, and underpins many applications in spectroscopy, photochemistry, and ultrafast science.

Previous works on electron density modeling often rely on graph-based neural networks, where atoms are treated as nodes and bonds or spatial cutoffs define edges. While powerful, this approach imposes an atomic representation that may be less natural for modeling the volumetric nature of the electron density in a continuous space. By contrast, volumetric representations preserve translational and rotational structure more directly, enable direct operator learning in function space, and can better capture delocalized electrons, charge density waves or excitations.

In this work, we propose a framework that combines a 3D convolutional autoencoder with a latent diffusion model (LDM) \cite{rombach2022high} to learn and evolve the electron density in a compressed latent space. The autoencoder first encodes volumetric electron density fields into a compact latent representation, preserving intrinsic spatial and physical structure while reducing dimensionality. The latent diffusion model then learns the conditional distribution of the next latent state given the current state, enabling autoregressive probabilistic generation of full electron density trajectories. Our formulation efficiently compresses the high-dimensional observed space into low-dimensional manifold while enabling the robust, long-horizon sampling without commonly seen drifting, collapse, or state stagnation problems.

\section{Generative Latent Space Dynamics of Electron Density}

\subsection{\textit{Ab-initio} Molecular Dynamics (AIMD)} 

\paragraph{Theory.} 

The time evolution of a many-electron system is, in principle, governed by the time-dependent many-body Schrödinger equation for all electrons and nuclei. In practice, this is computationally intractable beyond hundreds of atoms, and for weakly correlated systems the adiabatic Born–Oppenheimer (BO) approximation can be employed: the electronic and nuclear degrees of freedom are decoupled, and the nuclei move classically on the potential energy surface (PES) of the electronic ground state. For a fixed configuration of nuclei $\left\lbrace\mathbf{R}_I\right\rbrace$, the ground-state PES and \textit{electron number density} $\rho(\mathbf{r})$ are obtained from a self-consistent field (SCF) solution of the Kohn–Sham (KS) equations, \begin{gather}
    \left(-\frac{\hbar^2}{2m_e}\nabla^2 + V_\text{eff}(\mathbf{r})\right)\psi_i(\mathbf{r}) 
    = \varepsilon_i \psi_i(\mathbf{r}), \\
    \rho (\mathbf{r}) = \sum_i \left|\psi_i(\mathbf{r})\right|^2, \\ E = \min_{\rho} E\left[\rho\right],
\end{gather} where $V_\text{eff}(\mathbf{r})$ is the effective external potential parameterized by $\left\lbrace\mathbf{R}_I\right\rbrace$ and $E\left[\rho\right]$ is the energy functional of the electron density. After the SCF calculation is converged, the time evolution of atomic nuclei can be integrated from forces (and additional terms for ensembles other than microcanonical ensemble)
\begin{equation}
    M_I \ddot{\mathbf{R}}_I = -\nabla_{\mathbf{R}_I} E,
\end{equation} where $-\nabla_{\mathbf{R}_I} E$ is the Hellmann–Feynman force evaluated by the derivative of ground-state KS Hamiltonian and electron orbitals: $\mathbf{F}_I = - \Braket{\boldsymbol\psi_o|\frac{\partial\hat{H}_\text{KS}}{\partial \mathbf{R}_I}|\boldsymbol\psi_o}$. 

\paragraph{Dataset.} We generated a electron density trajectory of 32 Li atoms from isochoric-isothermal NVT AIMD simulation at \SI{800}{K}. For each ionic step, KS-DFT calculation is performed with generalized gradient approximation to search for the ground state electron density. Perdew-Burke-Ernzerhof (PBE) functional was used to describe exchange-correlation energy. The electron wave functions are expanded in plane-wave bases, with maximum energy cutoff \SI{680}{eV}. The AIMD trajectory was performed for \SI{10}{ps} at the timestep of \SI{2}{fs}, where the first \SI{8}{ps} was used for training and the last \SI{2}{ps} as test set. The AIMD and electron density trajectories were performed using GPAW \cite{mortensen2024gpaw} and ASE \cite{larsen2017atomic}. We recorded both total and pseudo electron densities for each frame. Nonetheless, only the pseudo electron density is variationally optimized and contains rich bonding information in the projected-augmented wave formalism \cite{blochl1994projector,mortensen2005real}. We use pseudo electron density as model learning objective and hereafter denote the pseudo electron number density as electron density throughout the work.

\begin{wrapfigure}{r}{0.3\textwidth}
        \includegraphics[width=\linewidth]{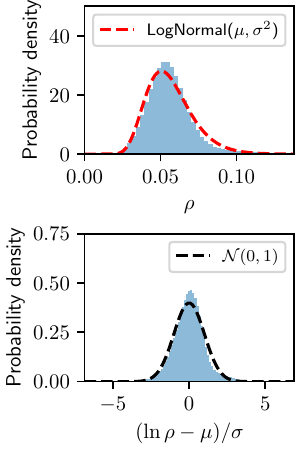}
    \caption{\textbf{Data distribution.} Pseudo electron density distribution of Li32 trajectory approximately follows single log-normal distribution with $\mu = -2.911, \sigma = 0.271$. Only training set is visualized.}
    \label{fig:rho}
\end{wrapfigure}

We analyzed the distribution of electron density in \Cref{fig:rho} and found that the value is roughly log-normally distributed for our toy system. We describe the observation below.

\begin{proposition}[Log-normal distribution of $\rho$]
The electron (number) density $\rho(\mathbf{r}, \tau)$ from an equilibrated AIMD trajectory is approximately log-normally distributed in the space-time dimensions \begin{equation}
    \ln \rho(\mathbf{r}, \tau) \sim \mathcal{N}\left(\mu, \sigma^2\right)
\end{equation}, where $\mu$ and $\sigma^2$ is the mean and variance of the Gaussian distribution.
\end{proposition}

To leverage this nice property, we therefore normalize our electron density data by logarithm and shift-scale transformation \begin{equation}
    \tilde{\rho} = \frac{\ln \rho - \mu}{\sigma},
    \label{eq:normalized}
\end{equation} where $\mu = -2.911$ and $\sigma = 0.271$ are the mean and standard deviation of $\ln \rho$ in training set. The probability density of $\tilde{\rho}(\mathbf{r}, \tau)$ is thereby close to standard normal distribution $\mathcal{N}(0, 1)$, as shown in the bottom panel of \Cref{fig:rho}. This transformation allows flexible reconstruction space for the decoder and diffusion models, and naturally enforces positivity with inverse relation $\exp\left(\sigma\tilde{\rho}+\mu\right) > 0$. 

\paragraph{State representation.} The state space of the classical atomistic system is determined by atomic positions and velocities. Similarly, in order to be complete, the state of the electron density should be at least described by both number density and its time derivative, as the time derivative of number density is linked to charge current density $\mathbf{j}(\mathbf{r}, \tau)$ by the continuity equation: $\dot{\rho}(\mathbf{r}, \tau) = -\nabla\mathbf{j}(\mathbf{r}, \tau)$. We formally specify each frame at the \textit{physical} time $\tau$ with the state as: \begin{multline}
    s(\tau) = \left(\mathbf{A}(\tau), \rho(\mathbf{r}, \tau), \dot{\rho}(\mathbf{r}, \tau)\right) \\ \in \mathbb{R}^{3 \times 3} \times \mathbb{R}_{\ge 0}^{N_1 \times N_2 \times N_3} \times \mathbb{R}^{N_1 \times N_2 \times N_3}, 
\end{multline} where $\mathbf{A} = \left[\mathbf{a}_1, \mathbf{a}_2, \mathbf{a}_3\right]$ is unit cell matrix of three lattice vectors, $\mathbf{R}, \dot{\mathbf{R}}$ are the ionic positions and velocities, and $N_i$ are the number of grid points along three dimensions. Since we have lifted the positivity constraint on electron density by \cref{eq:normalized} and fixed the cubic cell geometry under NVT ensemble, the state representation reduces to \begin{multline}
    s(\tau) = \left(L, \tilde{\rho}(\mathbf{r}, \tau), \dot{\tilde{\rho}}(\mathbf{r}, \tau)\right) \\ \in \mathbb{R} \times \mathbb{R}^{N_1 \times N_2 \times N_3} \times \mathbb{R}^{N_1 \times N_2 \times N_3}, 
\end{multline} where $L$ is the lattice constant (the length of cubic cell vector). We further leave $L$ out from the state representation in this work as the volume is fixed throughout the AIMD trajectory. In principle, the lattice parameters could be easily placed back by concatenation in the latent space.

\subsection{Latent Diffusion Model for 3D Scalar Field}

\begin{figure}
    \centering
    \includegraphics[width=0.9\linewidth]{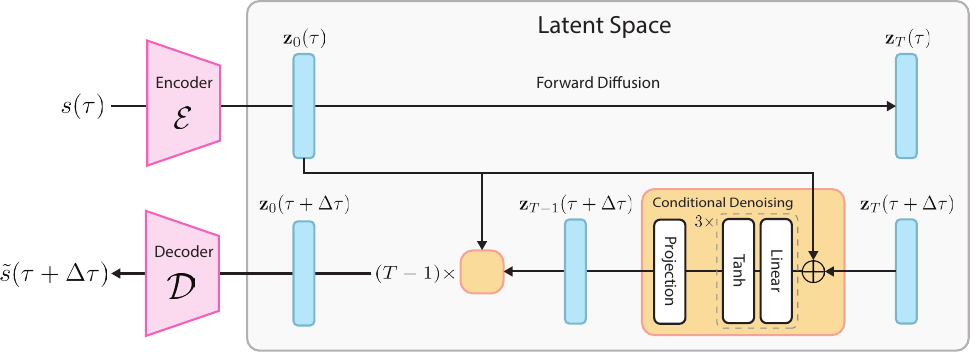}
    \caption{\textbf{Illustration of latent space generative process} to emulate the dynamics of electron density through the combination of autoencoder and latent diffusion model. At each physical time step $\tau$, the model conditions the latent denoiser with the current latent state $\mathbf{z}_0(\tau)$ to predict the next latent state $\mathbf{z}_0(\tau+\Delta\tau)$. The decoder maps the latent state back to physical space $\tilde{s}(\tau+\Delta\tau) = \mathcal{D}(\mathbf{z}_0(\tau+\Delta\tau))$.}
    \label{fig:arch}
\end{figure}

\paragraph{Model architecture.} Our model is conceptually inspired by latent diffusion model (LDM) \cite{rombach2022high}, but has undergone multiple major modifications to suit our spatiotemporal forecasting setting. Whereas the original LDM was designed for 2D image synthesis, inpainting, and related computer vision tasks, we generalize the learning task from 2D pixels to 3D voxel grids and recast the temporal prediction task as conditional probabilistic generation in the latent space (\Cref{fig:arch}).

The encoder $\mathcal{E}$ maps the initial physical state $s \in \mathbb{R}^{2 \times N_1 \times N_2 \times N_3}$ into a compact 1D latent vector $z = \mathcal{E}(s) \in \mathbb{R}^c$ via a sequence of 3D convolutional layers \cite{fukushima1980neocognitron, lecun2002gradient, krizhevsky2012imagenet}, residual blocks  \cite{he2015deepresiduallearningimage} with ELU activation \cite{clevert2015fast}, and circular padding to respect periodic boundaries. Downsampling is performed using strided convolutions, progressively increasing channel depth while reducing spatial resolution, followed by an average global pooling to produce a 1D fixed-size latent representation. The decoder $\mathcal{D}$ starts from a learned projection of the latent vector into a low-resolution 3D feature map, and then successively applies upsampling, residual blocks, and a final convolution to recover the voxel field $\tilde{s} = \mathcal{D}(z)$.

The diffusion process happens in the latent space. To not confuse diffusion timestep $t$ with physical time $\tau$, we denote forward and inverse diffusion steps as subscripts. In the forward diffusion process, Gaussian noise is gradually added to the sampled latents $\mathbf{z}_0$ in $T$ steps with linear variance schedule $\beta_{1:T}$, producing an array of corrupted samples $\mathbf{z}_{0:T}$: \begin{equation}
    q(\mathbf{z}_t \vert \mathbf{z}_{t-1}) = \mathcal{N}(\mathbf{z}_t; \sqrt{1 - \beta_t} \mathbf{z}_{t-1}, \beta_t\mathbf{I}), \quad
q(\mathbf{z}_{1:T} \vert \mathbf{z}_0) = \prod^T_{t=1} q(\mathbf{z}_t \vert \mathbf{z}_{t-1}).
\end{equation} 

The corrupted samples gradually lose their distinguishable features as diffusion step increases and approach Gaussian distribution. To recover the samples from the Gaussian, our goal is to learn a denoising predictor $p_\theta$ that approximates the conditional probabilities in the reverse diffusion process starting from a Gaussian at $p(\mathbf{z}_T) = \mathcal{N}(\mathbf{x}_T; \mathbf{0}, \mathbf{I})$: \begin{equation}
    p_\theta(\mathbf{z}_{0:T}) = p(\mathbf{z}_T) \prod^T_{t=1} p_\theta(\mathbf{z}_{t-1} \vert \mathbf{z}_t), \quad
p_\theta(\mathbf{z}_{t-1} \vert \mathbf{z}_t) = \mathcal{N}(\mathbf{z}_{t-1}; \boldsymbol{\mu}_\theta(\mathbf{z}_t, t), \boldsymbol{\Sigma}_\theta(\mathbf{z}_t, t)).
\end{equation} The learning objective \cite{ho2020denoising, rombach2022high} is to train a noise predictor $\boldsymbol{\epsilon}_\theta$ for the corrupted sample $\mathbf{z}_t = \mathcal{E}(\mathbf{s}_t)$ on the variation bound resembling denoising score matching \cite{song2019generative} \begin{equation}
    \mathcal{L}_\text{LDM} := \mathbb{E}_{t \sim [0, T], \mathcal{E}(\mathbf{s}_0), \boldsymbol{\epsilon}\sim \mathcal{N}(\mathbf{0}, \mathbf{I})} \left[\|\boldsymbol{\epsilon} - \boldsymbol{\epsilon}_\theta\left(\mathbf{z}_t, t\right)\|^2\right]
\end{equation} with step uniformly sampled $t \sim [1, T]$. As the latents have been compressed into 1D vectors, our denoising network $\boldsymbol{\epsilon}_\theta$ adopts a simple multi-layer perceptron (MLP) conditioned by time embeddings with sinusoidal positional encoding.

\paragraph{Conditional generation.} To recast a generative model into an autoregressive model, we condition the denoising network $\epsilon_\theta$ on the latent representation of the current state $\mathbf{z}(\tau) = \mathcal{E}(\mathbf{s}(\tau))$ to reconstruct the next state $\mathbf{z}(\tau + \Delta\tau)$. Formally, it is achieved by concatenating the current latent state $\mathbf{z}_0(\tau)$ in the input: \begin{equation}
    \mathcal{L}_\text{LDM} := \mathbb{E}_{t, \mathbf{z}_t(\tau), \mathbf{z}_t(\tau + \Delta \tau), \boldsymbol{\epsilon}} \left[\|\boldsymbol{\epsilon} - \boldsymbol{\epsilon}_\theta\left(\mathbf{z}_t(\tau + \Delta \tau), t, \mathbf{z}_0(\tau)\right)\|^2\right], 
\end{equation} where we denote physical time as $\tau$ and diffusion step as $t$ here. At the inference time, the model generates the next state by denoising from Gaussian distribution on the condition of current state $\mathbf{s}(\tau)$.

\paragraph{Loss and regularization.} To ensure that the statistical distribution of the generated electron density matches the ground truth, we incorporate a smooth, differentiable version of the Jensen-Shannon Divergence (JSD) as a loss term. Standard histograms are non-differentiable due to their discrete binning process, making them unsuitable for gradient-based optimization. To overcome this, we implement a \textit{soft histogram} where each data point contributes to multiple bins, weighted by a Gaussian kernel based on its proximity to the bin centers. This creates a smooth and differentiable approximation of probability distribution. We compute two such distributions, $P$ for the model's reconstruction and $Q$ for the target data, and then calculate the JSD between them. The JSD is a symmetric and smoothed measure of similarity between two probability distributions, defined as the average of the Kullback-Leibler (KL) divergences from each distribution to their midpoint average, providing a stable and robust loss signal for training.

The process is formalized by first calculating the soft probability distribution $P$ for a set of data points $\{x_i\}$ and bins with centers $\{b_j\}$: \begin{equation}
\pi_P(j) = \frac{\sum_i \exp\left(-\frac{(\text{clip}(x_i, b_{\min}, b_{\max}) - b_j)^2}{2\sigma^2}\right)}{\sum_k \sum_i \exp\left(-\frac{(\text{clip}(x_i, b_{\min}, b_{\max}) - b_k)^2}{2\sigma^2}\right)}.
\end{equation}

Inspired by the previous work on generalized JSD loss \cite{englesson2021generalized}, our empirical test shows that the scaled JSD by a constant factor $Z = -(1 - \pi_Q)\ln{(1-\pi_Q)}$ achieves better alignment between generated and target distributions. Given the distribution $P$ from the reconstructed data and $Q$ from the ground truth data, the scaled JSD loss $\mathcal{L}_\text{sJSD}$ is estimated from two probabilistic densities $\pi_P \sim P, \pi_Q \sim Q$: \begin{equation}
\mathcal{L}_{\mathrm{sJSD}} = \frac{1}{Z} D_{\mathrm{JS}}(P \,\|\, Q) = \frac{1}{2Z} \left[ D_{\mathrm{KL}}(P \,\|\, M) + D_{\mathrm{KL}}(Q \,\|\, M) \right], \quad\text{where} \quad M = \frac{P + Q}{2}.
\label{eq:sJSD}
\end{equation} 

The total training objective adds the additional reconstruction loss from CNN autoencoder $\mathcal{L}_\text{AE}$ and the denoising loss from LDM $\mathcal{L}_\text{LDM}$: \begin{equation}
    \mathcal{L} =  \mathcal{L}_\text{AE} + \lambda_1 \mathcal{L}_\text{LDM} + \lambda_2 \mathcal{L}_\text{sJSD}
    \label{eq:loss}, 
\end{equation} where $\lambda_1$ and $\lambda_2$ are tunable weights.

\section{Experiments}

We found that for electron density prediction, the ML model could still achieve low reconstruction error with average prediction on most of the grid points, as only a few of them have concentrated electron density. In such cases, the model learns only the fuzzy average of the input distribution and could not preserve the distributional attributes of the electron densities. Therefore, we apply sJSD loss \cref{eq:sJSD} on the reconstructed electron densities, with $\lambda_1=0.1$ and $\lambda_2=10$ in \cref{eq:loss} used. 

\begin{figure}[!t]
    \centering
    \includegraphics[width=\linewidth]{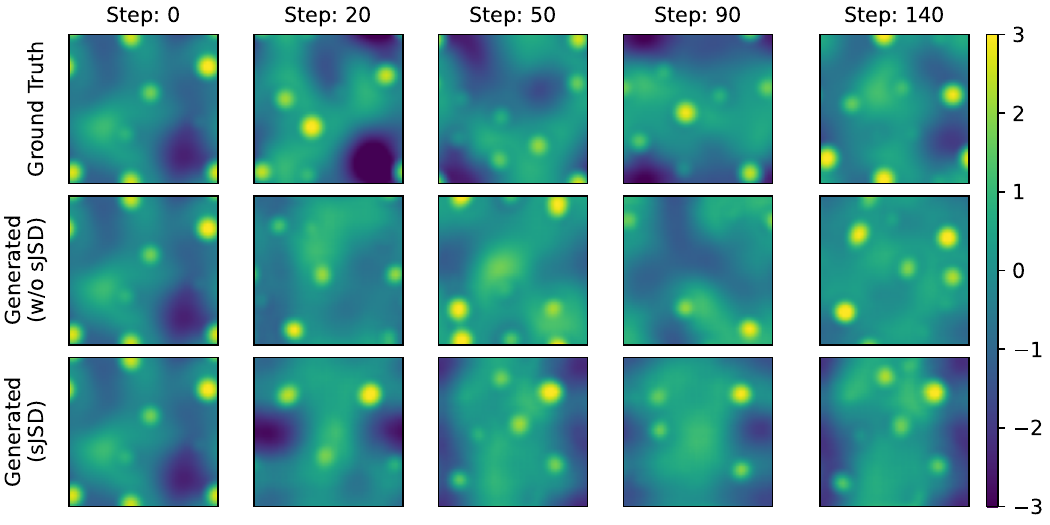}
    \caption{\textbf{Autoregressive trajectories.} LDM generates and evolves electron density qualitatively consistent with unseen test trajectory of Li atoms at \SI{800}{K}. The model trained with sJSD loss has less delocalized electron distribution similar to the ground truth. 2D slices at the middle plane along $x$ direction are shown. The grid values presented are normalized pseudo charge density $\tilde{\rho}$ (\cref{eq:normalized}, \cref{fig:rho}). See \Cref{fig:rho-dist-sq} for probability density and structure factor comparison.}
    \label{fig:forecast}
\end{figure}

\begin{figure}[!t]
    \centering
    \includegraphics{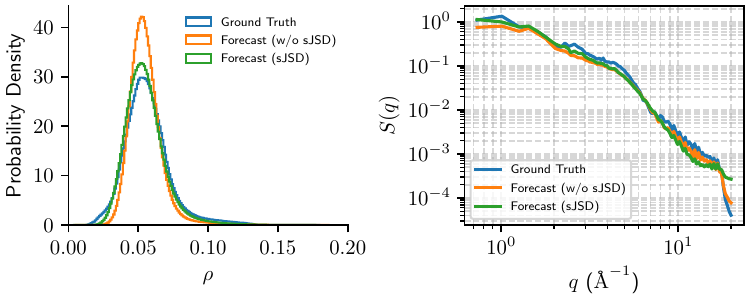}
    \caption{\textbf{Statistical consistency of generated trajectories.} Probability distribution and static structure factor $S(q)$ of test-set (ground-truth) and generated (forecast) electron densities. The LDM model trained with sJSD loss outperforms the model without sJSD. The static structure factor is obtained by averaging over the unrolled trajectory frames. The detailed $S(q)$ calculation can be found in \Cref{si:sq}.}
    \label{fig:rho-dist-sq}
\end{figure}

\Cref{fig:forecast} presents the generated autoregressive traejctories on two models, with and without sJSD losses. Generally, the model with sJSD regularization generates electron densities with visual characteristics closer to the ground-truth test trajectory. \Cref{fig:rho-dist-sq} further compares the distributional similarity between two generated and ground-truth, log-normal like distributions. The model trained with sJSD loss clearly demonstrates more similar distribution to test trajectory than the one without sJSD regularization, which overemphasizes the population around \SI{0.05}{\angstrom^{-3}}. See \Cref{fig:ae-mse} for training loss comparison.

To ensure the model also generates correct spatial distribution as well, we further compare the structure factor $S(q)$ of the three trajectories in \Cref{fig:rho-dist-sq}. It can be seen that convolutional autoencoder, while highly constrained in design, is able to robustly preserve the correct spatial correlation as the dynamics evolves. The model with sJSD loss aligns better with ground truth in low $q$, long wavelength regime, demonstrating better ability to capture long-range spatial correlation of electron density. This is, however, at the expense of the high $q$, short wavelength regime, but the difference is orders of magnitude smaller. The kink near $q = 4$ is consistent with the atomic radius of a lithium atom ($\approx \SI{1.52}{\angstrom}$): $$q = \frac{2\pi}{a} = 4, \quad a \approx 1.57.$$

\section{Related Work}

\paragraph{Electron density prediction.} Early ML approaches to electron density prediction focused on mapping molecular geometries to ground-state densities using kernel methods or feedforward networks. More recent works leverage equivariant graph neural networks (GNNs) \cite{koker2024higher, jorgensen2022equivariant, jorgensen2020deepdft} or symmetry-preserving neural architectures \cite{fu2024recipe, koker2024higher} to capture both local and long-range correlations. However, these methods typically predict static, ground-state densities, with limited exploration of dynamical or time-dependent behavior. 

\paragraph{Dimensionality reduction.} Compressing high-dimensional electron density data into a lower-dimensional latent space could potentially facilitate efficient learning and simulation. The possible techniques range from classical approaches (\textit{e.g.} principal component analysis, proper orthogonal decomposition \cite{kim2021efficient, mcbane2021component, choi2020gradient, choi2019space, copeland2022reduced, choi2021space, choi2020sns, lauzon2024s, cheung2023local}, multi-dimensional scaling \cite{davison2000multidimensional}) to more modern, non-linear approaches (\textit{e.g.} autoencoder \cite{kim2022fast, diaz2024fast, fries2022lasdi, he2023glasdi, bonneville2024gplasdi, tran2024weak, park2024tlasdi, he2025physics, kadeethum2022non}, kernal PCA, Isomap \cite{tenenbaum2000global}) to manifold learning \cite{meilua2024manifold}.

\paragraph{Operator learning.} Neural operator frameworks, such as the Fourier Neural Operator (FNO) \cite{li2020fourier}, Factorized Fourier Neural Operators \cite{tran2021factorized}, DeepONet \cite{lu2019deeponet}, and among others, have emerged as powerful tools for learning solution operators to partial differential equations (PDEs). Physics-informed operator learning has also been applied to dynamical fields, with methods like DISCO \cite{morel2025disco} enabling spatiotemporal prediction from sparse observations. Relatedly, Koopman operator theory \cite{koopman1931hamiltonian} offers foundation for obtaining the linear embedding for nonlinear dynamical systems. Recently, diffusion models have been also adopted to learn the dynamics of molecular simulations \cite{hsu2024score} and PDE solutions \cite{rozet2025lost}.

\paragraph{Neural ODE/PDE solvers.} Neural Ordinary Differential Equation (Neural ODE) and Partial Differential Equation (Neural PDE) solvers leverage deep learning to model governing equations in function space, bypassing the need for explicit discretization. Neural ODEs, introduced by \citet{chen2018neural}, parameterize the derivative of a system with a neural network and integrate it using numerical ODE solvers, enabling adaptive time-stepping and memory-efficient training via adjoint methods. Neural PDE solvers extend this idea to spatially extended systems, often combining neural operators or physics-informed neural networks (PINNs) with domain knowledge to learn solutions across space and time \cite{raissi2019physics,cuomo2022scientific,fries2022lasdi}. These approaches are particularly powerful for modeling high-dimensional, nonlinear, or multiscale systems where traditional solvers may be computationally expensive, offering mesh-free generalization and the ability to learn from sparse or noisy data.

\section{Discussion}

\paragraph{Learning in Fourier space.} We found that Fourier representation of electron density, while efficient in dimension reduction, struggles to compress meaningful latent space representation due to increased difficulty to handle translation symmetry and ambiguity of multiple equivalent phase shifts. Our preliminary experiments revealed that naïve MLP autoencoder is prone to overfit to Fourier representation and has poor extrapolation capability. While FNO has demonstrated strong success in weather forecasting and PDE problems \cite{li2020fourier, kurth2023fourcastnet}, our preliminary test on FNO shows pronounced drift and quickly becomes unstable during trajectory rollout. In contrast, the average pooling bottleneck at the end of our encoder enforces the translation invariance for latent representation and enables our model to be stable without noticeable shift. The future investigation of models that are equivariant to phase shift and robust to translation will be important.

\paragraph{Another bitter lesson.} In our experiments, the direct enforcement of physics-informed neural network (PINN) loss fails to learn the useful latent representations to reliably roll out the dynamics. The most straightforward autoencoder (AE) trained against PINN loss on charge conservation and spatial gradients is found easily overfitting to the training distribution. When the trajectory enters the unseen region in the latent space of AE, the decoder either starts to have significant drift in the output space or freezes in the inactive regions or fixed points caused by the ``dead neurons''. We show that small CNN model with average pooling bottleneck and sJSD loss, although having less parameters than MLP AE, transformer, neural operator and others widely used for PDE solving, could generalize better to unseen AIMD trajectory and reliably roll out the system state without drifting or freezing in the latent space during test time. This is counterintuitive to many recent efforts in PDE solving with more scalable models like transformers and neural operators, arguably because the expressive model tends to overfit to the noise of high-dimensional data ($48^3 = 110,592$ for each frame in our case), and the the underlying quantum mechanical nature of the system is far more complicated and discontinuous than continuum problem routinely solved.

\paragraph{Limitations and opportunities.} Since ions move classically in the current problem setting, our current experiments can in fact be bootstrapped by running machine learning interatomic potential \cite{batatia_foundation_2023} and electron density prediction model in alternate steps. However, we see that our framework could be extended to even more challenging calculations such as time-dependent density functional theory (TDDFT) \cite{runge1984density} and equation-of-motion coupled-cluster theory (EOM-CC) \cite{stanton1993equation}, where the data is even more scarce and the time-dependent electron density is the important data for computing diverse properties. Currently, our framework awaits generalization to multiple atomic species to facilitate the usage for actual DFT calculations for either initial guess or direction optimization of electron density in the conceptually similar manner as orbital-free DFT (OF-DFT) and their variants \cite{mi2023orbital, ligneres2005introduction, jiang2021time}. A complementary direction is the integration of this line of work into the recently proposed data-driven finite element method (DD-FEM) framework \cite{choi2025defining}. Since the generation of high-fidelity electron density data from quantum molecular dynamics or time-dependent density functional theory is computationally expensive, DD-FEM offers a path to bridge this gap. By learning reduced representations and operators in a small subdomain or element scale, DD-FEM could enable scalable surrogate modeling of density fields, lowering the cost of data generation while retaining physical interpretability. Such an approach has the potential to extend the utility of generative latent dynamics models toward practical, multiscale applications in spectroscopy, photochemistry, and ultrafast science.

\section{Acknowledgment}
This work was performed under the auspices of the U.S. Department of Energy (DOE), by Lawrence Livermore National Laboratory (LLNL) under Contract No. DE-AC52–07NA27344 and was supported by Laboratory Directed Research and Development funding under projects 24-ERD-035. All the authors appreciate the fruitful discussion with Dr. Jean-Luc Fattebert from Oak Ridge National Laboratory. Yuan Chiang thanks Prof. Aditi Krishnapriyan, Yiheng Du, Xiao Liu, and Chang-Han Chen for fruitful discussion, and Prof. Mark Asta for his support and guidance as Yuan's PhD advisor in UC Berkeley. Yuan Chiang also appreciates the financial support from Taiwan-UC Berkeley Fellowship, LBNL, and LLNL. LLNL release number: LLNL-PROC-2010268.

\bibliographystyle{unsrtnat}
\bibliography{references}

\begin{thebibliography}{62}
\providecommand{\natexlab}[1]{#1}
\providecommand{\url}[1]{\texttt{#1}}
\expandafter\ifx\csname urlstyle\endcsname\relax
  \providecommand{\doi}[1]{doi: #1}\else
  \providecommand{\doi}{doi: \begingroup \urlstyle{rm}\Url}\fi

\bibitem[Gubler et~al.(2025)Gubler, Sch{\"a}fer, Behler, and Goedecker]{gubler2025accuracy}
Moritz Gubler, Moritz~R Sch{\"a}fer, J{\"o}rg Behler, and Stefan Goedecker.
\newblock Accuracy of charge densities in electronic structure calculations.
\newblock \emph{The Journal of Chemical Physics}, 162\penalty0 (9), 2025.

\bibitem[Pulay(1980)]{pulay1980convergence}
P{\'e}ter Pulay.
\newblock Convergence acceleration of iterative sequences. the case of scf iteration.
\newblock \emph{Chemical physics letters}, 73\penalty0 (2):\penalty0 393--398, 1980.

\bibitem[Broyden(1965)]{broyden1965class}
Charles~G Broyden.
\newblock A class of methods for solving nonlinear simultaneous equations.
\newblock \emph{Mathematics of computation}, 19\penalty0 (92):\penalty0 577--593, 1965.

\bibitem[Das and Gavini(2023)]{das2023accelerating}
Sambit Das and Vikram Gavini.
\newblock Accelerating self-consistent field iterations in kohn-sham density functional theory using a low-rank approximation of the dielectric matrix.
\newblock \emph{Physical Review B}, 107\penalty0 (12):\penalty0 125133, 2023.

\bibitem[Fu et~al.(2024)Fu, Rosen, Bystrom, Wang, Musaelian, Kozinsky, Smidt, and Jaakkola]{fu2024recipe}
Xiang Fu, Andrew Rosen, Kyle Bystrom, Rui Wang, Albert Musaelian, Boris Kozinsky, Tess Smidt, and Tommi Jaakkola.
\newblock A recipe for charge density prediction.
\newblock \emph{Advances in Neural Information Processing Systems}, 37:\penalty0 9727--9752, 2024.

\bibitem[Koker et~al.(2024)Koker, Quigley, Taw, Tibbetts, and Li]{koker2024higher}
Teddy Koker, Keegan Quigley, Eric Taw, Kevin Tibbetts, and Lin Li.
\newblock Higher-order equivariant neural networks for charge density prediction in materials.
\newblock \emph{npj Computational Materials}, 10\penalty0 (1):\penalty0 161, 2024.

\bibitem[J{\o}rgensen and Bhowmik(2022)]{jorgensen2022equivariant}
Peter~Bj{\o}rn J{\o}rgensen and Arghya Bhowmik.
\newblock Equivariant graph neural networks for fast electron density estimation of molecules, liquids, and solids.
\newblock \emph{npj Computational Materials}, 8\penalty0 (1):\penalty0 183, 2022.

\bibitem[J{\o}rgensen and Bhowmik(2020)]{jorgensen2020deepdft}
Peter~Bj{\o}rn J{\o}rgensen and Arghya Bhowmik.
\newblock Deepdft: Neural message passing network for accurate charge density prediction.
\newblock \emph{arXiv preprint arXiv:2011.03346}, 2020.

\bibitem[Achar et~al.(2023)Achar, Bernasconi, and Johnson]{achar2023machine}
Siddarth~K Achar, Leonardo Bernasconi, and J~Karl Johnson.
\newblock Machine learning electron density prediction using weighted smooth overlap of atomic positions.
\newblock \emph{Nanomaterials}, 13\penalty0 (12):\penalty0 1853, 2023.

\bibitem[Li et~al.(2025)Li, Sharir, Yuan, and Chan]{li2025image}
Chenghan Li, Or~Sharir, Shunyue Yuan, and Garnet Kin-Lic Chan.
\newblock Image super-resolution inspired electron density prediction.
\newblock \emph{Nature Communications}, 16\penalty0 (1):\penalty0 4811, 2025.

\bibitem[Rombach et~al.(2022)Rombach, Blattmann, Lorenz, Esser, and Ommer]{rombach2022high}
Robin Rombach, Andreas Blattmann, Dominik Lorenz, Patrick Esser, and Bj{\"o}rn Ommer.
\newblock High-resolution image synthesis with latent diffusion models.
\newblock In \emph{Proceedings of the IEEE/CVF conference on computer vision and pattern recognition}, pages 10684--10695, 2022.

\bibitem[Mortensen et~al.(2024)Mortensen, Larsen, Kuisma, Ivanov, Taghizadeh, Peterson, Haldar, Dohn, Sch{\"a}fer, J{\'o}nsson, et~al.]{mortensen2024gpaw}
Jens~J{\o}rgen Mortensen, Ask~Hjorth Larsen, Mikael Kuisma, Aleksei~V Ivanov, Alireza Taghizadeh, Andrew Peterson, Anubhab Haldar, Asmus~Ougaard Dohn, Christian Sch{\"a}fer, Elvar~{\"O}rn J{\'o}nsson, et~al.
\newblock Gpaw: An open python package for electronic structure calculations.
\newblock \emph{The Journal of Chemical Physics}, 160\penalty0 (9), 2024.

\bibitem[Larsen et~al.(2017)Larsen, Mortensen, Blomqvist, Castelli, Christensen, Du{\l}ak, Friis, Groves, Hammer, Hargus, et~al.]{larsen2017atomic}
Ask~Hjorth Larsen, Jens~J{\o}rgen Mortensen, Jakob Blomqvist, Ivano~E Castelli, Rune Christensen, Marcin Du{\l}ak, Jesper Friis, Michael~N Groves, Bj{\o}rk Hammer, Cory Hargus, et~al.
\newblock The atomic simulation environment—a python library for working with atoms.
\newblock \emph{Journal of Physics: Condensed Matter}, 29\penalty0 (27):\penalty0 273002, 2017.

\bibitem[Bl{\"o}chl(1994)]{blochl1994projector}
Peter~E Bl{\"o}chl.
\newblock Projector augmented-wave method.
\newblock \emph{Physical review B}, 50\penalty0 (24):\penalty0 17953, 1994.

\bibitem[Mortensen et~al.(2005)Mortensen, Hansen, and Jacobsen]{mortensen2005real}
Jens~J{\o}rgen Mortensen, Lars~Bruno Hansen, and Karsten~Wedel Jacobsen.
\newblock Real-space grid implementation of the projector augmented wave method.
\newblock \emph{Physical Review B—Condensed Matter and Materials Physics}, 71\penalty0 (3):\penalty0 035109, 2005.

\bibitem[Fukushima(1980)]{fukushima1980neocognitron}
Kunihiko Fukushima.
\newblock Neocognitron: A self-organizing neural network model for a mechanism of pattern recognition unaffected by shift in position.
\newblock \emph{Biological cybernetics}, 36\penalty0 (4):\penalty0 193--202, 1980.

\bibitem[LeCun et~al.(2002)LeCun, Bottou, Bengio, and Haffner]{lecun2002gradient}
Yann LeCun, L{\'e}on Bottou, Yoshua Bengio, and Patrick Haffner.
\newblock Gradient-based learning applied to document recognition.
\newblock \emph{Proceedings of the IEEE}, 86\penalty0 (11):\penalty0 2278--2324, 2002.

\bibitem[Krizhevsky et~al.(2012)Krizhevsky, Sutskever, and Hinton]{krizhevsky2012imagenet}
Alex Krizhevsky, Ilya Sutskever, and Geoffrey~E Hinton.
\newblock Imagenet classification with deep convolutional neural networks.
\newblock \emph{Advances in neural information processing systems}, 25, 2012.

\bibitem[He et~al.(2015)He, Zhang, Ren, and Sun]{he2015deepresiduallearningimage}
Kaiming He, Xiangyu Zhang, Shaoqing Ren, and Jian Sun.
\newblock Deep residual learning for image recognition, 2015.
\newblock URL \url{https://arxiv.org/abs/1512.03385}.

\bibitem[Clevert et~al.(2015)Clevert, Unterthiner, and Hochreiter]{clevert2015fast}
Djork-Arn{\'e} Clevert, Thomas Unterthiner, and Sepp Hochreiter.
\newblock Fast and accurate deep network learning by exponential linear units (elus).
\newblock \emph{arXiv preprint arXiv:1511.07289}, 4\penalty0 (5):\penalty0 11, 2015.

\bibitem[Ho et~al.(2020)Ho, Jain, and Abbeel]{ho2020denoising}
Jonathan Ho, Ajay Jain, and Pieter Abbeel.
\newblock Denoising diffusion probabilistic models.
\newblock \emph{Advances in neural information processing systems}, 33:\penalty0 6840--6851, 2020.

\bibitem[Song and Ermon(2019)]{song2019generative}
Yang Song and Stefano Ermon.
\newblock Generative modeling by estimating gradients of the data distribution.
\newblock \emph{Advances in neural information processing systems}, 32, 2019.

\bibitem[Englesson and Azizpour(2021)]{englesson2021generalized}
Erik Englesson and Hossein Azizpour.
\newblock Generalized jensen-shannon divergence loss for learning with noisy labels.
\newblock \emph{Advances in Neural Information Processing Systems}, 34:\penalty0 30284--30297, 2021.

\bibitem[Kim et~al.(2021)Kim, Wang, and Choi]{kim2021efficient}
Youngkyu Kim, Karen Wang, and Youngsoo Choi.
\newblock Efficient space--time reduced order model for linear dynamical systems in python using less than 120 lines of code.
\newblock \emph{Mathematics}, 9\penalty0 (14):\penalty0 1690, 2021.

\bibitem[McBane and Choi(2021)]{mcbane2021component}
Sean McBane and Youngsoo Choi.
\newblock Component-wise reduced order model lattice-type structure design.
\newblock \emph{Computer methods in applied mechanics and engineering}, 381:\penalty0 113813, 2021.

\bibitem[Choi et~al.(2020{\natexlab{a}})Choi, Boncoraglio, Anderson, Amsallem, and Farhat]{choi2020gradient}
Youngsoo Choi, Gabriele Boncoraglio, Spenser Anderson, David Amsallem, and Charbel Farhat.
\newblock Gradient-based constrained optimization using a database of linear reduced-order models.
\newblock \emph{Journal of Computational Physics}, 423:\penalty0 109787, 2020{\natexlab{a}}.

\bibitem[Choi and Carlberg(2019)]{choi2019space}
Youngsoo Choi and Kevin Carlberg.
\newblock Space--time least-squares petrov--galerkin projection for nonlinear model reduction.
\newblock \emph{SIAM Journal on Scientific Computing}, 41\penalty0 (1):\penalty0 A26--A58, 2019.

\bibitem[Copeland et~al.(2022)Copeland, Cheung, Huynh, and Choi]{copeland2022reduced}
Dylan~Matthew Copeland, Siu~Wun Cheung, Kevin Huynh, and Youngsoo Choi.
\newblock Reduced order models for lagrangian hydrodynamics.
\newblock \emph{Computer Methods in Applied Mechanics and Engineering}, 388:\penalty0 114259, 2022.

\bibitem[Choi et~al.(2021)Choi, Brown, Arrighi, Anderson, and Huynh]{choi2021space}
Youngsoo Choi, Peter Brown, William Arrighi, Robert Anderson, and Kevin Huynh.
\newblock Space--time reduced order model for large-scale linear dynamical systems with application to boltzmann transport problems.
\newblock \emph{Journal of Computational Physics}, 424:\penalty0 109845, 2021.

\bibitem[Choi et~al.(2020{\natexlab{b}})Choi, Coombs, and Anderson]{choi2020sns}
Youngsoo Choi, Deshawn Coombs, and Robert Anderson.
\newblock Sns: A solution-based nonlinear subspace method for time-dependent model order reduction.
\newblock \emph{SIAM Journal on Scientific Computing}, 42\penalty0 (2):\penalty0 A1116--A1146, 2020{\natexlab{b}}.

\bibitem[Lauzon et~al.(2024)Lauzon, Cheung, Shin, Choi, Copeland, and Huynh]{lauzon2024s}
Jessica~T Lauzon, Siu~Wun Cheung, Yeonjong Shin, Youngsoo Choi, Dylan~M Copeland, and Kevin Huynh.
\newblock S-opt: A points selection algorithm for hyper-reduction in reduced order models.
\newblock \emph{SIAM Journal on Scientific Computing}, 46\penalty0 (4):\penalty0 B474--B501, 2024.

\bibitem[Cheung et~al.(2023)Cheung, Choi, Copeland, and Huynh]{cheung2023local}
Siu~Wun Cheung, Youngsoo Choi, Dylan~Matthew Copeland, and Kevin Huynh.
\newblock Local lagrangian reduced-order modeling for the rayleigh-taylor instability by solution manifold decomposition.
\newblock \emph{Journal of Computational Physics}, 472:\penalty0 111655, 2023.

\bibitem[Davison and Sireci(2000)]{davison2000multidimensional}
Mark~L Davison and Stephen~G Sireci.
\newblock Multidimensional scaling.
\newblock In \emph{Handbook of applied multivariate statistics and mathematical modeling}, pages 323--352. Elsevier, 2000.

\bibitem[Kim et~al.(2022)Kim, Choi, Widemann, and Zohdi]{kim2022fast}
Youngkyu Kim, Youngsoo Choi, David Widemann, and Tarek Zohdi.
\newblock A fast and accurate physics-informed neural network reduced order model with shallow masked autoencoder.
\newblock \emph{Journal of Computational Physics}, 451:\penalty0 110841, 2022.

\bibitem[Diaz et~al.(2024)Diaz, Choi, and Heinkenschloss]{diaz2024fast}
Alejandro~N Diaz, Youngsoo Choi, and Matthias Heinkenschloss.
\newblock A fast and accurate domain decomposition nonlinear manifold reduced order model.
\newblock \emph{Computer Methods in Applied Mechanics and Engineering}, 425:\penalty0 116943, 2024.

\bibitem[Fries et~al.(2022)Fries, He, and Choi]{fries2022lasdi}
William~D Fries, Xiaolong He, and Youngsoo Choi.
\newblock Lasdi: Parametric latent space dynamics identification.
\newblock \emph{Computer Methods in Applied Mechanics and Engineering}, 399:\penalty0 115436, 2022.

\bibitem[He et~al.(2023)He, Choi, Fries, Belof, and Chen]{he2023glasdi}
Xiaolong He, Youngsoo Choi, William~D Fries, Jonathan~L Belof, and Jiun-Shyan Chen.
\newblock glasdi: Parametric physics-informed greedy latent space dynamics identification.
\newblock \emph{Journal of Computational Physics}, 489:\penalty0 112267, 2023.

\bibitem[Bonneville et~al.(2024)Bonneville, Choi, Ghosh, and Belof]{bonneville2024gplasdi}
Christophe Bonneville, Youngsoo Choi, Debojyoti Ghosh, and Jonathan~L Belof.
\newblock Gplasdi: Gaussian process-based interpretable latent space dynamics identification through deep autoencoder.
\newblock \emph{Computer Methods in Applied Mechanics and Engineering}, 418:\penalty0 116535, 2024.

\bibitem[Tran et~al.(2024)Tran, He, Messenger, Choi, and Bortz]{tran2024weak}
April Tran, Xiaolong He, Daniel~A Messenger, Youngsoo Choi, and David~M Bortz.
\newblock Weak-form latent space dynamics identification.
\newblock \emph{Computer Methods in Applied Mechanics and Engineering}, 427:\penalty0 116998, 2024.

\bibitem[Park et~al.(2024)Park, Cheung, Choi, and Shin]{park2024tlasdi}
Jun Sur~Richard Park, Siu~Wun Cheung, Youngsoo Choi, and Yeonjong Shin.
\newblock tlasdi: Thermodynamics-informed latent space dynamics identification.
\newblock \emph{Computer Methods in Applied Mechanics and Engineering}, 429:\penalty0 117144, 2024.

\bibitem[He et~al.(2025)He, Tran, Bortz, and Choi]{he2025physics}
Xiaolong He, April Tran, David~M Bortz, and Youngsoo Choi.
\newblock Physics-informed active learning with simultaneous weak-form latent space dynamics identification.
\newblock \emph{International Journal for Numerical Methods in Engineering}, 126\penalty0 (1):\penalty0 e7634, 2025.

\bibitem[Kadeethum et~al.(2022)Kadeethum, Ballarin, Choi, O’Malley, Yoon, and Bouklas]{kadeethum2022non}
Teeratorn Kadeethum, Francesco Ballarin, Youngsoo Choi, Daniel O’Malley, Hongkyu Yoon, and Nikolaos Bouklas.
\newblock Non-intrusive reduced order modeling of natural convection in porous media using convolutional autoencoders: comparison with linear subspace techniques.
\newblock \emph{Advances in Water Resources}, 160:\penalty0 104098, 2022.

\bibitem[Tenenbaum et~al.(2000)Tenenbaum, Silva, and Langford]{tenenbaum2000global}
Joshua~B Tenenbaum, Vin~de Silva, and John~C Langford.
\newblock A global geometric framework for nonlinear dimensionality reduction.
\newblock \emph{science}, 290\penalty0 (5500):\penalty0 2319--2323, 2000.

\bibitem[Meil{\u{a}} and Zhang(2024)]{meilua2024manifold}
Marina Meil{\u{a}} and Hanyu Zhang.
\newblock Manifold learning: What, how, and why.
\newblock \emph{Annual Review of Statistics and Its Application}, 11\penalty0 (1):\penalty0 393--417, 2024.

\bibitem[Li et~al.(2020)Li, Kovachki, Azizzadenesheli, Liu, Bhattacharya, Stuart, and Anandkumar]{li2020fourier}
Zongyi Li, Nikola Kovachki, Kamyar Azizzadenesheli, Burigede Liu, Kaushik Bhattacharya, Andrew Stuart, and Anima Anandkumar.
\newblock Fourier neural operator for parametric partial differential equations.
\newblock \emph{arXiv preprint arXiv:2010.08895}, 2020.

\bibitem[Tran et~al.(2021)Tran, Mathews, Xie, and Ong]{tran2021factorized}
Alasdair Tran, Alexander Mathews, Lexing Xie, and Cheng~Soon Ong.
\newblock Factorized fourier neural operators.
\newblock \emph{arXiv preprint arXiv:2111.13802}, 2021.

\bibitem[Lu et~al.(2019)Lu, Jin, and Karniadakis]{lu2019deeponet}
Lu~Lu, Pengzhan Jin, and George~Em Karniadakis.
\newblock Deeponet: Learning nonlinear operators for identifying differential equations based on the universal approximation theorem of operators.
\newblock \emph{arXiv preprint arXiv:1910.03193}, 2019.

\bibitem[Morel et~al.(2025)Morel, Han, and Oyallon]{morel2025disco}
Rudy Morel, Jiequn Han, and Edouard Oyallon.
\newblock Disco: learning to discover an evolution operator for multi-physics-agnostic prediction.
\newblock \emph{arXiv preprint arXiv:2504.19496}, 2025.

\bibitem[Koopman(1931)]{koopman1931hamiltonian}
Bernard~O Koopman.
\newblock Hamiltonian systems and transformation in hilbert space.
\newblock \emph{Proceedings of the National Academy of Sciences}, 17\penalty0 (5):\penalty0 315--318, 1931.

\bibitem[Hsu et~al.(2024)Hsu, Sadigh, Bulatov, and Zhou]{hsu2024score}
Tim Hsu, Babak Sadigh, Vasily Bulatov, and Fei Zhou.
\newblock Score dynamics: Scaling molecular dynamics with picoseconds time steps via conditional diffusion model.
\newblock \emph{Journal of Chemical Theory and Computation}, 20\penalty0 (6):\penalty0 2335--2348, 2024.

\bibitem[Rozet et~al.(2025)Rozet, Ohana, McCabe, Louppe, Lanusse, and Ho]{rozet2025lost}
Fran{\c{c}}ois Rozet, Ruben Ohana, Michael McCabe, Gilles Louppe, Fran{\c{c}}ois Lanusse, and Shirley Ho.
\newblock Lost in latent space: An empirical study of latent diffusion models for physics emulation.
\newblock \emph{arXiv preprint arXiv:2507.02608}, 2025.

\bibitem[Chen et~al.(2018)Chen, Rubanova, Bettencourt, and Duvenaud]{chen2018neural}
Ricky~TQ Chen, Yulia Rubanova, Jesse Bettencourt, and David~K Duvenaud.
\newblock Neural ordinary differential equations.
\newblock \emph{Advances in neural information processing systems}, 31, 2018.

\bibitem[Raissi et~al.(2019)Raissi, Perdikaris, and Karniadakis]{raissi2019physics}
Maziar Raissi, Paris Perdikaris, and George~E Karniadakis.
\newblock Physics-informed neural networks: A deep learning framework for solving forward and inverse problems involving nonlinear partial differential equations.
\newblock \emph{Journal of Computational physics}, 378:\penalty0 686--707, 2019.

\bibitem[Cuomo et~al.(2022)Cuomo, Di~Cola, Giampaolo, Rozza, Raissi, and Piccialli]{cuomo2022scientific}
Salvatore Cuomo, Vincenzo~Schiano Di~Cola, Fabio Giampaolo, Gianluigi Rozza, Maziar Raissi, and Francesco Piccialli.
\newblock Scientific machine learning through physics--informed neural networks: Where we are and what’s next.
\newblock \emph{Journal of Scientific Computing}, 92\penalty0 (3):\penalty0 88, 2022.

\bibitem[Kurth et~al.(2023)Kurth, Subramanian, Harrington, Pathak, Mardani, Hall, Miele, Kashinath, and Anandkumar]{kurth2023fourcastnet}
Thorsten Kurth, Shashank Subramanian, Peter Harrington, Jaideep Pathak, Morteza Mardani, David Hall, Andrea Miele, Karthik Kashinath, and Anima Anandkumar.
\newblock Fourcastnet: Accelerating global high-resolution weather forecasting using adaptive fourier neural operators.
\newblock In \emph{Proceedings of the platform for advanced scientific computing conference}, pages 1--11, 2023.

\bibitem[Batatia et~al.(2023)Batatia, Benner, Chiang, Elena, Kovács, Riebesell, Advincula, Asta, Baldwin, Bernstein, Bhowmik, Blau, Cărare, Darby, De, Della~Pia, Deringer, Elijošius, El-Machachi, Fako, Ferrari, Genreith-Schriever, George, Goodall, Grey, Han, Handley, Heenen, Hermansson, Holm, Jaafar, Hofmann, Jakob, Jung, Kapil, Kaplan, Karimitari, Kroupa, Kullgren, Kuner, Kuryla, Liepuoniute, Margraf, Magdău, Michaelides, Moore, Naik, Niblett, Norwood, O'Neill, Ortner, Persson, Reuter, Rosen, Schaaf, Schran, Sivonxay, Stenczel, Svahn, Sutton, van~der Oord, Varga-Umbrich, Vegge, Vondrák, Wang, Witt, Zills, and Csányi]{batatia_foundation_2023}
Ilyes Batatia, Philipp Benner, Yuan Chiang, Alin~M. Elena, Dávid~P. Kovács, Janosh Riebesell, Xavier~R. Advincula, Mark Asta, William~J. Baldwin, Noam Bernstein, Arghya Bhowmik, Samuel~M. Blau, Vlad Cărare, James~P. Darby, Sandip De, Flaviano Della~Pia, Volker~L. Deringer, Rokas Elijošius, Zakariya El-Machachi, Edvin Fako, Andrea~C. Ferrari, Annalena Genreith-Schriever, Janine George, Rhys E.~A. Goodall, Clare~P. Grey, Shuang Han, Will Handley, Hendrik~H. Heenen, Kersti Hermansson, Christian Holm, Jad Jaafar, Stephan Hofmann, Konstantin~S. Jakob, Hyunwook Jung, Venkat Kapil, Aaron~D. Kaplan, Nima Karimitari, Namu Kroupa, Jolla Kullgren, Matthew~C. Kuner, Domantas Kuryla, Guoda Liepuoniute, Johannes~T. Margraf, Ioan-Bogdan Magdău, Angelos Michaelides, J.~Harry Moore, Aakash~A. Naik, Samuel~P. Niblett, Sam~Walton Norwood, Niamh O'Neill, Christoph Ortner, Kristin~A. Persson, Karsten Reuter, Andrew~S. Rosen, Lars~L. Schaaf, Christoph Schran, Eric Sivonxay, Tamás~K. Stenczel, Viktor Svahn, Christopher
  Sutton, Cas van~der Oord, Eszter Varga-Umbrich, Tejs Vegge, Martin Vondrák, Yangshuai Wang, William~C. Witt, Fabian Zills, and Gábor Csányi.
\newblock A foundation model for atomistic materials chemistry, December 2023.
\newblock URL \url{http://arxiv.org/abs/2401.00096}.
\newblock arXiv:2401.00096 [cond-mat, physics:physics].

\bibitem[Runge and Gross(1984)]{runge1984density}
Erich Runge and Eberhard~KU Gross.
\newblock Density-functional theory for time-dependent systems.
\newblock \emph{Physical review letters}, 52\penalty0 (12):\penalty0 997, 1984.

\bibitem[Stanton and Bartlett(1993)]{stanton1993equation}
John~F Stanton and Rodney~J Bartlett.
\newblock The equation of motion coupled-cluster method. a systematic biorthogonal approach to molecular excitation energies, transition probabilities, and excited state properties.
\newblock \emph{The Journal of chemical physics}, 98\penalty0 (9):\penalty0 7029--7039, 1993.

\bibitem[Mi et~al.(2023)Mi, Luo, Trickey, and Pavanello]{mi2023orbital}
Wenhui Mi, Kai Luo, SB~Trickey, and Michele Pavanello.
\newblock Orbital-free density functional theory: An attractive electronic structure method for large-scale first-principles simulations.
\newblock \emph{Chemical Reviews}, 123\penalty0 (21):\penalty0 12039--12104, 2023.

\bibitem[Lign{\`e}res and Carter(2005)]{ligneres2005introduction}
Vincent~L Lign{\`e}res and Emily~A Carter.
\newblock An introduction to orbital-free density functional theory.
\newblock In \emph{Handbook of materials modeling: methods}, pages 137--148. Springer, 2005.

\bibitem[Jiang and Pavanello(2021)]{jiang2021time}
Kaili Jiang and Michele Pavanello.
\newblock Time-dependent orbital-free density functional theory: Background and pauli kernel approximations.
\newblock \emph{Physical Review B}, 103\penalty0 (24):\penalty0 245102, 2021.

\bibitem[Choi et~al.(2025)Choi, Cheung, Kim, Tsai, Diaz, Zanardi, Chung, Copeland, Kendrick, Anderson, et~al.]{choi2025defining}
Youngsoo Choi, Siu~Wun Cheung, Youngkyu Kim, Ping-Hsuan Tsai, Alejandro~N Diaz, Ivan Zanardi, Seung~Whan Chung, Dylan~Matthew Copeland, Coleman Kendrick, William Anderson, et~al.
\newblock Defining foundation models for computational science: A call for clarity and rigor.
\newblock \emph{arXiv preprint arXiv:2505.22904}, 2025.

\end{thebibliography}

\newpage
\appendix
\setcounter{figure}{0}  
\setcounter{table}{0}   
\setcounter{equation}{0}  
\renewcommand{\thefigure}{S\arabic{figure}}
\renewcommand{\thetable}{S\arabic{table}}
\renewcommand{\theequation}{S\arabic{equation}}

\section{Structure factor $S(q)$ calculation}
\label{si:sq}

We compute the complex structure factor directly from the electron density field 
\(\rho(\mathbf{r})\) defined on a uniform grid with spacings 
\(\Delta x, \Delta y, \Delta z\) and voxel volume \(\Delta V=\Delta x\,\Delta y\,\Delta z\).
Using the FFT convention, the forward and inverse relations are
\begin{equation}
F(\mathbf{q}) \;=\; \sum_{\mathbf{r}} \rho(\mathbf{r})\,e^{-i\mathbf{q}\cdot\mathbf{r}}\,\Delta V,
\qquad
\rho(\mathbf{r}) \;=\; \frac{1}{V}\sum_{\mathbf{q}} F(\mathbf{q})\,e^{+i\mathbf{q}\cdot\mathbf{r}},
\end{equation}
where \(V=N_xN_yN_z\,\Delta V\) is the total volume. In our implementation we compute
\(F(\mathbf{q})\) via an FFT of \(\rho(\mathbf{r})\) and multiply by \(\Delta V\) to be
consistent with the continuous-transform normalization. We then form the \emph{amplitude}
spectrum (not intensity) by taking the complex modulus,
\begin{equation}
|F(\mathbf{q})| \;=\; \big|\mathcal{F}\{\rho(\mathbf{r})\}\big|.
\end{equation}
To obtain a rotationally invariant one-dimensional curve, we spherically average this
amplitude over shells of constant \(q=|\mathbf{q}|\):
\begin{equation}
S(q) \;\equiv\; \left\langle |F(\mathbf{q})| \right\rangle_{|\mathbf{q}|=q}
\;=\;
\frac{1}{N(q)}\sum_{\mathbf{q}\in[q-\Delta q/2,\,q+\Delta q/2]} |F(\mathbf{q})|,
\end{equation}
where \(N(q)\) is the number of reciprocal-grid points in the shell. In practice,
reciprocal vectors are constructed from FFT frequencies as
\(
q_x=2\pi n_x/L_x,\;
q_y=2\pi n_y/L_y,\;
q_z=2\pi n_z/L_z
\)
with \(L_\alpha = N_\alpha \Delta \alpha\). The \(\mathbf{q}=\mathbf{0}\) mode
is set to zero to remove the mean density contribution.

\begin{figure}[!h]
    \centering
    \includegraphics[width=0.5\linewidth]{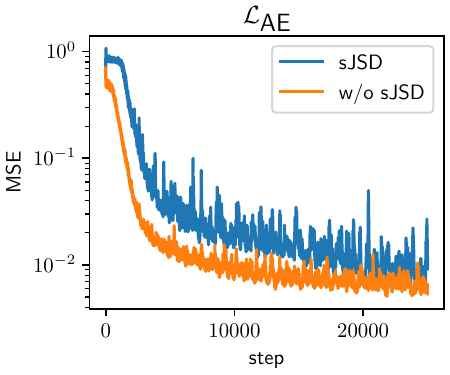}
    \caption{Mean square error (MSE) of autoencoder reconstruction for training set at each optimization step.}
    \label{fig:ae-mse}
\end{figure}

\end{document}